\documentclass{KapProc} 
\kluwerbib
\let\lcitebracket(
\let\rcitebracket)
\usepackage[dvips]{graphics}
\newcommand{\etal}{{et~al.}}
\newcommand{\lsim}{\raise0.3ex\hbox{$<$}\kern-0.75em{\lower0.65ex\hbox{$\sim$}}}
\newcommand{\gsim}{\raise0.3ex\hbox{$>$}\kern-0.75em{\lower0.65ex\hbox{$\sim$}}}
\begin{document}


\articletitle{The host galaxies of radio-loud and radio-quiet quasars}

\author{James S. Dunlop} 
\affil{
  Institute for Astronomy, Royal Observatory, Edinburgh EH9 3HJ, U.K.\\}

\begin{abstract}
\noindent
I review our knowledge of the properties of the host galaxies of radio-loud 
and radio-quiet quasars, both in comparison to each other and in the context 
of the general galaxy population. It is now clear that the hosts of radio-loud 
{\it and} radio-quiet quasars with $M_V < -23.5$ are virtually all 
massive elliptical galaxies.
The masses of these spheroids are as expected given the relationship between 
black-hole and spheroid mass found for nearby quiescent galaxies, as is the 
growing prevalence of disc components in the hosts of progressively fainter 
AGN. There is also now compelling evidence that quasar hosts are practically
indistinguishable from normal ellipticals, both in their basic structural 
parameters and in the old age of their dominant stellar populations; at low 
$z$ the nuclear activity is {\it not} associated with the formation of 
a significant fraction of the host galaxy.
While the long-held view that quasar radio power might be a simple function
of host morphology is now dead and buried, I argue that host-galaxy studies 
may yet play a crucial role in resolving the long-standing problem of the 
origin of radio loudness. Specifically there is growing evidence that 
radio-loud objects are powered by more massive black holes accreting at lower 
efficiency than their radio-quiet couterparts of comparable optical output.
A black-hole mass $> 10^9 {\rm M_{\odot}}$ appears to be a necessary (although 
perhaps not sufficient) condition for the production of radio jets of 
sufficient power to produce an FRII radio source within a massive galaxy halo.
\end{abstract}


\section{Introduction}

Studies of the host galaxies of low-redshift quasars are of crucial importance 
for defining the subset of the present-day galaxy population which is capable 
of producing quasar-level nuclear activity. They are also of value for 
constraining physical models of quasar evolution, 
for exploring the extent to which radio-loudness might be connected
with host-galaxy properties, and as a means to estimate the masses of the
central black holes which power the active nuclei.

Our view of low-redshift quasar hosts has been clarified enormously 
over the last five years, primarily due to the angular resolution and dynamic
range offered by the Hubble Space Telescope. In this overview I have therefore
chosen to concentrate on the results of recent, primarily HST-based studies 
of low-redshift quasars, and will only briefly mention the latest results at 
higher redshift which are discussed in detail elsewhere in these proceedings.
I have also chosen to centre the discussion around our own, recently-completed,
HST imaging study of the hosts of quasars at $z \simeq 0.2$. Preliminary
results from this programme can be found in McLure et al. (1999) and final 
results from the completed samples are presented by Dunlop et al. (2001). Here
I focus on a few of the main results from this study and discuss 
the extent to which other authors do or do not agree with our findings.

\section{Host galaxy luminosity, morphology and size}

After some initial confusion (e.g. Bahcall et al. 1994), recent HST-based
studies have now reached agreement that the hosts of all luminous quasars 
($M_V < -23.5$) are bright galaxies with $L > L^{\star}$ (Mclure et al. 1999,
McLeod \& McLeod 2001, Dunlop et al. 2001). However, it can be argued,
(with some justification) that this much had already been established from 
earlier ground-based studies (e.g. Taylor et al. 1996). 

In fact the major 
advance offered by the HST for the study of quasar hosts is that it has 
enabled host luminosity profiles to be measured over sufficient angular and 
dynamic range to allow a de Vaucouleurs $r^{1/4}$-law spheroidal component to 
be clearly distinguished from an exponential disc, at least for redshifts 
$z < 0.5$. In our own study 
this is the reason that we have been able to establish unambiguously 
that, at low $z$,  the hosts of both radio-loud quasars (RLQs) {\it and} 
radio-quiet quasars (RQQs) are undoubtedly 
massive ellipticals with (except for one RQQ in our sample) 
negligible disc components 
(McLure et al. 1999, Dunlop et al. 2001).
This result is illustrated in figure 1. 

\begin{figure}[h]
\vspace{10.5cm}
\centering
\setlength{\unitlength}{1mm}
\includegraphics{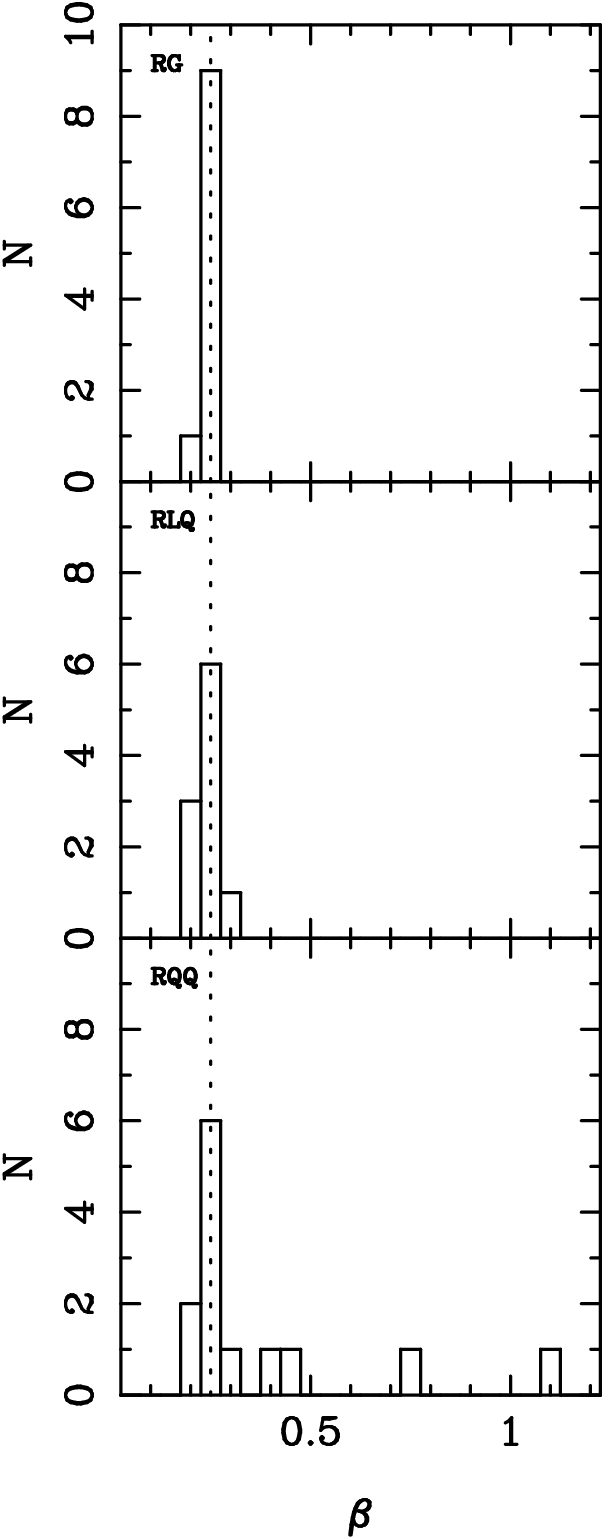}
\caption{Histograms of the best-fit values of $\beta$, where host-galaxy
surface brightness is proportional to $exp(-(r)^{\beta})$, shown for the
radio-galaxy, radio-loud quasar and radio-quiet quasar sub-samples imaged
with the HST by Dunlop et al. (2001). The dotted line at $\beta=0.25$
indicates a perfect de Vaucouleurs law, and all of the radio-loud hosts
are consistent with this within the errors. Two of the three RQQs with
hosts for which $\beta > 0.4$ transpire to be the two least luminous 
nuclei in the sample, and should really be reclassified as Seyferts.}
\label{betahist}
\end{figure}

Figure 1 confirms that the hosts of radio-loud quasars and radio galaxies all
follow essentially perfect de Vaucouleurs profiles, in good agreement with 
the results of other studies. The perhaps more surprising aspect of figure
1 is the extent to which our radio-quiet quasar sample is also dominated 
by spheroidal hosts. At first sight this might seem at odds with the results
of some other recent studies, such as those of Bahcall et al. (1997) and 
Hamilton et al. (2001) who report that 
approximately one third to one half of radio-quiet quasars lie in disc-dominated hosts.
However, on closer examination it becomes clear that there is no real 
contradiction provided one compares quasars of similar power. 
Specifically, if attention is confined to quasars with nuclear magnitudes 
$M_V < -23.5$ we find that 10 out of the 11 RQQs in our sample lie in ellipticals,
Bahcall et al. find that 6 of their 7 similarly-luminous quasars lie in 
ellipticals, while an examination of the data in Hamilton et al. shows that in 
fact at least 17 out of the 20 comparably-luminous RQQs in their archival sample 
also appear to lie in spheroidal hosts. 

\begin{figure}[h]
\vspace{6cm}
\centering
\setlength{\unitlength}{1mm}
\includegraphics{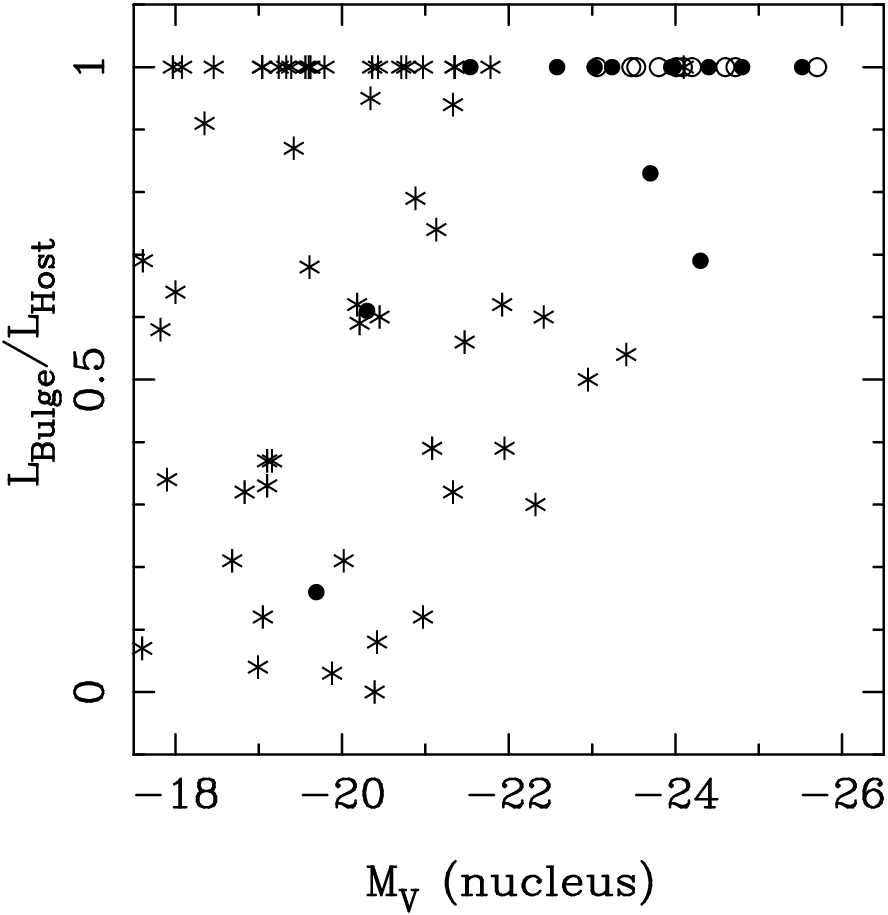}
\caption{The relative contribution of the spheroidal
component to the total luminosity of the host galaxy plotted against absolute
$V$-band
luminosity of the nuclear component. The plot shows the results
for our own HST sample (RLQs as open circles, RQQs as filled circles)
along with the results from Schade et al. (2000) for a larger sample of
X-ray selected AGN spanning a wider but lower range of 
optical luminosities (asterisks). 
This plot illustrates very clearly how disc-dominated
host galaxies become increasingly rare with increasing nuclear power, as
is expected if more luminous AGN are powered by more massive black holes
which, in turn, are housed in more massive spheroids.}
\end{figure}

It is thus now clear that above a given luminosity threshold we enter a regime
in which AGN can only be hosted by massive spheroids, regardless 
of radio power. It is also clear that, within the radio-quiet
population, significant disc components become more common at lower
nuclear luminosities. This dependence of host-galaxy morphology on nuclear 
luminosity is nicely demonstrated by combining our own results with those of 
Schade et al. (2000) who have studied the host galaxies of lower-luminosity
X-ray selected AGN. This I have done in figure 2 where the ratio
of bulge to total host luminosity is plotted as a function of nuclear 
optical 
power. Figure 2 is at least qualitatively as expected if black-hole mass is 
proportional to spheroid mass (Magorrian et al. 1998, Merritt \& Ferrarese 2001), and black-hole 
masses $> 5 \times 10^8
{\rm M_{\odot}}$ are required to produce quasars with $M_R < -23.5$.

In concluding this discussion of host morphology I should note that
there is at least some (albeit yet tentative) evidence that the hosts
of some of the most luminous quasars may in fact have a significant disc 
contribution
(Percival et al. 2001). At first sight this would appear to be at odds with
the appealingly simple picture presented in figure 2, and it will certainly
be interesting to see if this result survives the scrutiny of 
HST imaging currently underway. However, if confirmed, such a result 
need not contradict the universality of elliptical hosts, but rather might
mean that some of the most luminous quasars arise from the merger of the 
elliptical host
with a massive gas-rich disc galaxy, in which case the underlying massive elliptical
might (at least temporarily) appear to have acquired a significant disc 
component.

In our HST study we have also been able to break the well-known degeneracy
between host galaxy surface-brightness and size. This point is illustrated
by the fact that we have, for the first time, been able to demonstrate
that the hosts of RLQs and RQQs follow a Kormendy relation (figure 3). 
Moreover the  
slope ($2.90 \pm 0.2$) and normalization of this relation are identical to that 
displayed by normal quiescent massive ellipticals.
The average half-light radii of the host galaxies in our sub-samples are
11 kpc for the RGs, 12 kpc for the RLQs, and 8 kpc for the RQQs
($H_{0}=50$~km s$^{-1}$ Mpc$^{-1}$,
$\Omega_{m}=1.0$). For comparison the average
half-light radius of brightest-cluster galaxies observed by Schneider et
al. (1983) is 13 kpc.

\begin{figure}[h]

\vspace{6cm}
\centering
\setlength{\unitlength}{1mm}
\includegraphics{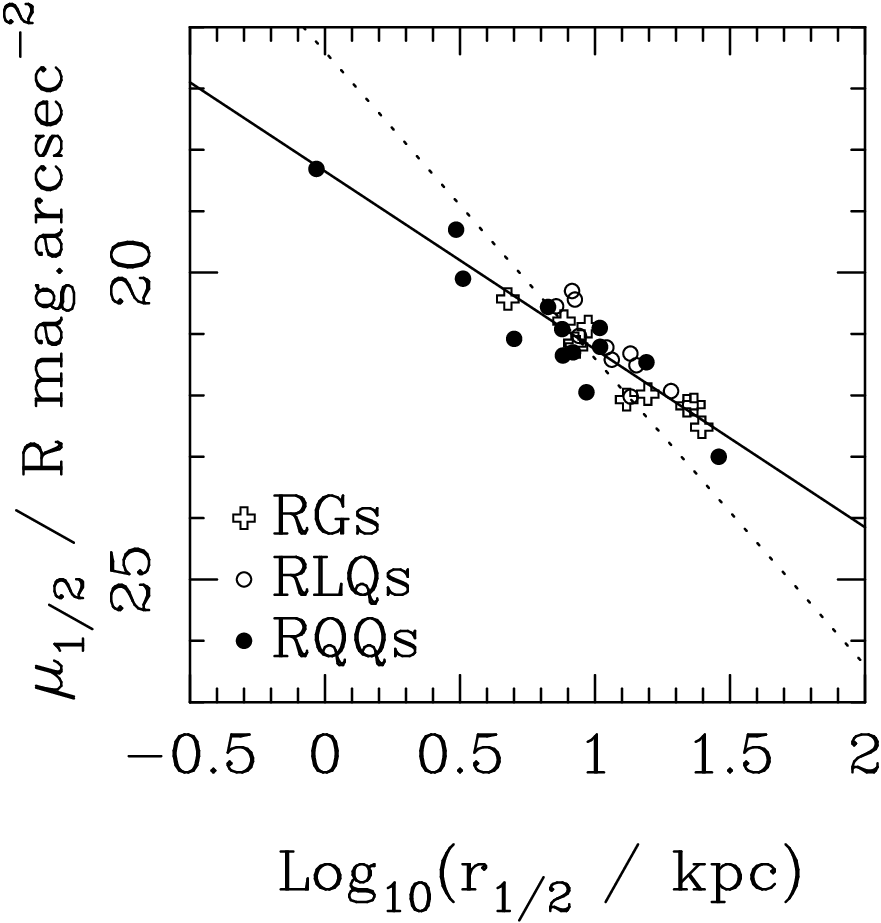}
\caption{The Kormendy relation followed by the hosts of all 33 powerful AGN
studied by Dunlop et al. (2001) with the HST. The solid line is the least-squares fit to the data
which has a slope of $2.90 \pm 0.2$, in excellent agreement with the slope of
2.95 found by Kormendy (1977) for inactive ellipticals.
For the few RQQs which have a disc
component the best-fitting bulge component has been plotted. }
\label{kormendy}
\end{figure}

\section{Host galaxy ages}

It is well known from simulations that the merger of two disc galaxies can 
produce  a remnant which displays a luminosity profile not dissimilar
to a de Vaucouleurs $r^{1/4}$-law. This raises the possibility that the 
apparently spheroidal nature of the quasar hosts discussed above might be
the result of a recent major merger which could also be responsible for 
stimulating the onset of nuclear activity. This would also be the natural
prediction of suggested evolutionary schemes in which ULIRGs are 
presumed to be the precursors of RQQs. Could a recent merger of two massive
gas-rich discs be simultaneously responsible for the triggering of 
nuclear activity and the production of an apparently spheroidal host?

The answer appears to be no. One piece of evidence against such a picture
comes from the fact that, as mentioned above, the Kormendy relation
displayed by quasar hosts appears to be indistinguishable from that
of quiescent, well-evolved massive ellipticals. However, a more 
direct test comes from attempts to determine the ages of the dominant stellar
populations in the quasar hosts. Within our own sample we have attempted to 
estimate the ages of the host galaxies both from optical-infrared colours
(now possible for the first time by combining our HST images with our
pre-existing UKIRT data; Taylor et al. 1996) and from deep optical
off-nuclear spectroscopy (Nolan et al. 2000). The results
of this investigation are summarized in figure 4, which shows that the 
hosts of both radio-loud and radio-quiet quasars are dominated by old
well-evolved stellar populations (with typically less than 1\% of stellar
mass involved in recent star-formation activity).
There are currently no comparably-extensive studies of 
host-galaxy stellar populations with this result can be compared. However,
Canalizo \& Stockton (2000) have published results from a more detailed 
spectroscopic study of three objects, one of which, Mkn 1014,
is also in our RQQ sample. This is in fact the only quasar host
for which we have found clear spectrosocpic evidence of A-star 
features and a significant (albeit still only $\simeq 2$\% by mass) 
young stellar population. It 
is presumably no coincidence that this is also the only quasar in our sample
which was detected by IRAS, and the only host which displays spectacular
tidal-tail features comparable to those commonly found in images of 
ULIRGs (see Sanders, this proceedings). However, even for this apparently
star-forming quasar host, Canalizo \& Stockton agree that $\simeq 95$\% of the host is dominated
by an old well-evolved stellar population (although they argue that
$5-8$\% of the galaxy has been involved in recent star formation).

\begin{figure}[h]
\vspace{7cm}
\centering
\setlength{\unitlength}{1mm}
\includegraphics{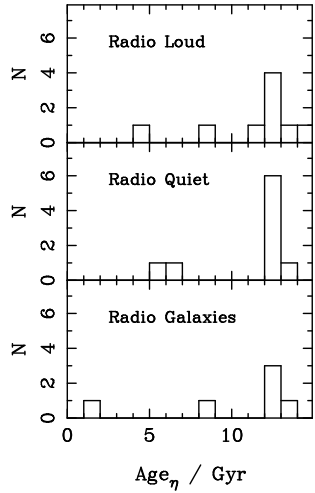}
\caption{The age distribution of the dominant stellar populations in the
sub-samples of host galaxies studied by Nolan et al. (2001). The ages
were derived by fitting a 3-component model (comprising scattered quasar
light, a young (0.1Gyr) stellar population, and an underlying stellar
population of age ranging from 0.1 to 14 Gyr) simultaneously to
off-nuclear optical spectra and the $R-K$ colours of the host galaxies.
The dominant populations in the hosts of both radio-loud and radio-quiet
AGN are predominantly old (12-14 Gyr) as is found for quiescent
elliptical galaxies.
}
\end{figure}

In summary, at least for low-redshift quasars, the timescale of the 
{\it primary} star-formation epoch in the host appears to be completely disconnected
from that of the more recent nuclear activity which has resulted in the object
featuring in quasar catalogues. The production of a low-redshift quasar 
only seems to require the massive, 
well-evolved spheroid housing the massive black hole to undergo
a relatively minor interaction. In contrast the 
production of a ULIRG seems to require a major merger between two massive 
galaxies at least one of which is gas rich. Present evidence suggests that the 
overlap between these two phenomena is rather limited at low redshift, and that 
the ULIRG $\rightarrow$ quasar evolutionary route can only apply 
to a fairly small subset of objects (e.g. Mkn 1014). Of course at high 
redshift the prospect for star-formation and nuclear activity having 
completely disconnected timescales is much more limited, and it seems likely 
that the first epoch of quasar activity in a massive galaxy is closely 
connected with massive (possibly dust-enshrouded) star-formation activity 
in the host (e.g. Fabian 1999, Archibald et al. 2001).
 
\section{Black hole masses and radio loudness}

Having established that the hosts of quasars are massive spheroids one can 
estimate the mass of  
their central black holes using the 
relationship between spheroid luminosity and black-hole mass recently 
derived from dynamical studies of nearby galaxies (e.g. Magorrian et al. 
1998, Merritt \& Ferrarese 2001). While undoubtedly uncertain to within a factor
of a few, the attractiveness of this approach is that it allows an estimate
of the central black-hole mass which is independent of any of the 
observed properties of the active nucleus. This estimate can then be compared
with, for example, an estimate based on the assumption
that the nucleus is accreting at the Eddington limit.

Using one of the most recent determinations of the black-hole:spheroid 
mass relationship, $M_{bh} = 0.0013 M_{spheroid}$ (Merritt \& Ferrarese 2001),
we find average black-hole mass estimates of $\langle M_{bh} \rangle =
1.5 \times 10^9 {\rm M_{\odot}}$ for the RLQs in our sample,
and $\langle M_{bh} \rangle = 0.9 \times 10^9 {\rm M_{\odot}}$ for the 
RQQs. This subtle but apparently persistent difference (see below) arises
directly from the fact that, although perfectly matched in optical {\it
nuclear}
luminosity, the hosts of our RQQs are, on average, $\simeq 1.5 - 2$ 
times less luminous than the hosts of their 
radio-loud counterparts.

\begin{figure}[h]
\vspace{6cm}
\centering
\setlength{\unitlength}{1mm}
\includegraphics{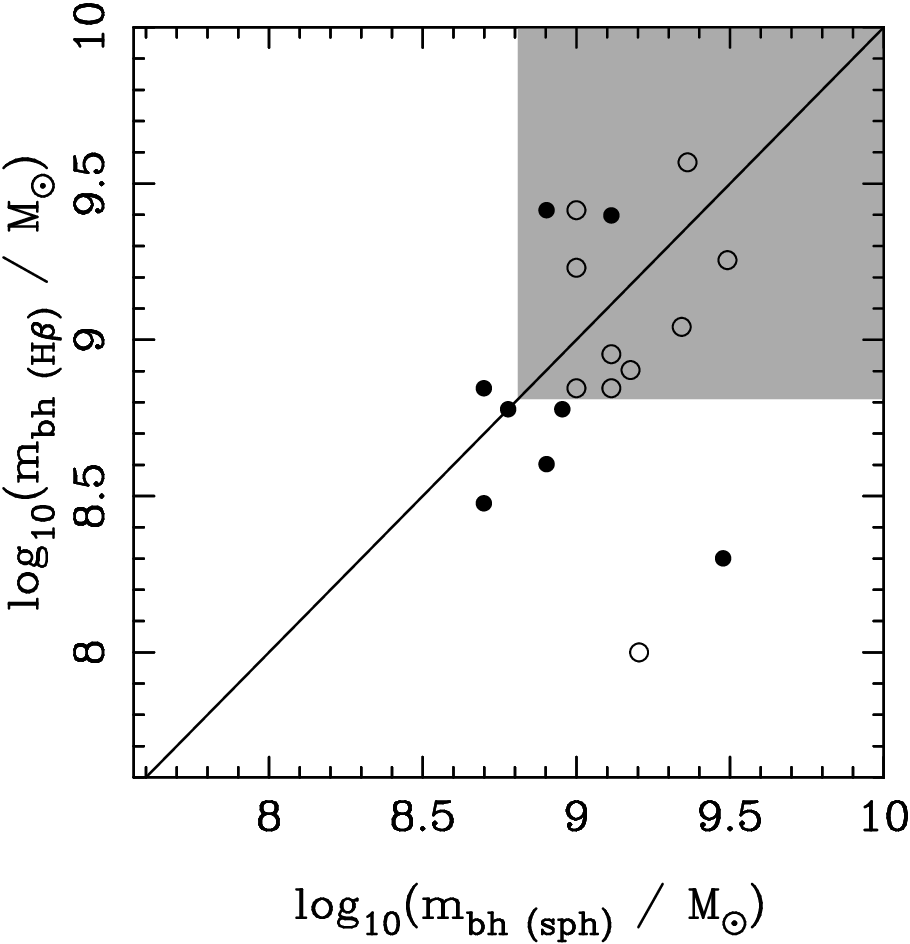}
\caption{A comparison between the black-hole masses of quasars as predicted from
host-galaxy spheroidal luminosity by Dunlop et al. (2001), and the corresponding
values determined from $H\beta$ line-width by McLure \& Dunlop (2001).
The shaded area is shown to demonstrate that there is a region in which
both approaches agree that $M_{bh} \gsim 10^9 M_{\odot}$, and that this
region contains all except one of the RLQs (open circles), but excludes all 
except 2 of the RQQs (filled circles).}
\label{bhcorr2}
\end{figure}

A comparison of the resulting predicted Eddington luminosities with 
the actual observed output of the quasar nuclei leads to the conclusion that 
most of the RLQs in our sample are emitting at $\simeq 5 - 10$\% 
of their potential Eddington limit, while the radio-quiet objects span 
a wider range in efficiency, from
$\simeq 10$\% to 100\% of the Eddington limit.

\begin{figure}[h]
\vspace{4.4cm}
\centering
\setlength{\unitlength}{1mm}
\includegraphics{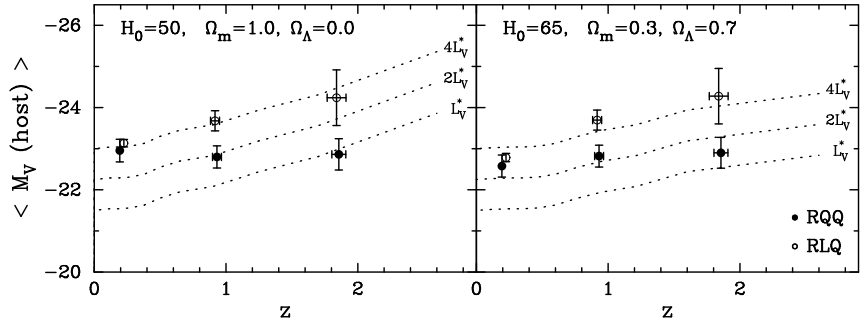}
\caption{
Mean absolute $V$-band magnitude versus mean redshift for the
host galaxies of the RLQs (open circles) and RQQs (filled circles) in
the NICMOS study of Kukula et al. (2001). Also shown is the subset of 5 RLQs and 7
RQQs from the Dunlop et al. (2001) WFPC2 study of quasars at $z\sim0.2$ which have total
(host $+$ nuclear) luminosities in the same range as the high-redshift
samples ($-24\geq M_{V}\geq-25$). Error bars show the standard error
on the mean. The dotted lines show the luminosity evolution of present
day $L^{\star}$, $2L^{\star}$ and $4L^{\star}$ elliptical galaxies,
assuming a formation epoch of $z=5$ with a single rapid burst of
starformation followed by passive evolution thereafter. LH panel:
assuming $H_{0}=50$~km s$^{-1}$ Mpc$^{-1}$,
$\Omega_{m}=1.0$ and $\Omega_{\Lambda}=0.0$. RH panel:
$H_{0}=65$~km s$^{-1}$ Mpc$^{-1}$, $\Omega_{m}=0.3$ and
$\Omega_{\Lambda}=0.7$.}
\end{figure}

The above black-hole mass estimates can also be compared with values 
derived, completely independently, from an analysis of the velocity 
width of the $H_{\beta}$ lines in the quasar nuclear spectra under the
assumption that the broad-line region is gravitationally bound. This has been
a growth industry in recent years (e.g. Wandel 1999, Laor 2000), bolstered
by estimates of the size of the broad-line region from
reverberation mapping of Seyfert galaxies. Recently Ross McLure and 
I have applied this technique to estimate the masses of the black holes
which power the quasars we have imaged with the HST. The results
are remarkably similar to the values described above, with 
the $H_{\beta}$ line-width yielding $\langle M_{bh}\rangle = 1 \times 10^9 
{\rm M_{\odot}}$ for the RLQs, and $\langle M_{bh} \rangle
= 5 \times 10^8 {\rm M_{\odot}}$ for the RQQs. 

Such agreement (to within
a factor of two - figure 5) suggests that these mass estimates should be 
taken seriously,
and of special interest is the fact that the apparent mass offset between
the black holes which power radio-loud and radio-quiet objects persists
(figure 5).
Indeed, given the uncertainties involved, the division in mass appears fairly
clean, at least in the sense that the radio-loud objects 
all lie above a certain mass threshold. Black-hole mass estimation 
from host spheroid luminosity leads to the 
conclusion that 9 out of the 10 RLQs have $M_{bh} > 10^9 {\rm M_{\odot}}$
while only 4 out of the 11 RQQs lie above this threshold.
From the $H_{\beta}$ analysis 11 out of 13 RLQs have $M_{bh} >
10^{8.8} {\rm M_{\odot}}$, 
while only 4 out of 17 RQQs lie in this regime (see McLure, these
proceedings). A similar conclusion has recently been reached by Laor
(2000).

There are a number of possible explanations for this apparent 
black-hole mass difference between radio-loud and radio-quiet objects.
Interestingly Blandford (2000) argues that highly-collimated 
jets might only be produced by sub-Eddington accretion. Thus it 
may simply be the case that by selecting 
RLQs and RQQs of comparable optical output, we are
guaranteed to find sub-Eddington accreters in the radio-loud sample, whereas
the radio-quiet sample can contain at least some less massive holes 
emitting at close to maximum efficiency. 

\section{The connection to high redshift}

The effective study of quasar hosts at high redshift is still in its
infancy. However, already it is becoming clear that the mass offset
between RQQ and RLQ hosts described above appears to grow with increasing
redshift (see figure 6, plus Kukula et al. (2001), and contributions from
Kukula, Ridgway, Impey and Rix in these proceedings), lending additional
credence to its reality. Specifically, for
the same nuclear luminosity, RQQ hosts at $z \simeq 2$ appear to be a
factor of 2-3 less massive than either their low-z counterparts or their
$z \simeq 2$ radio-loud counterparts. It is too early to say whether this
is due to changes in the host population, or simply due to (on average)
more efficient black-hole fueling revealing more clearly the mass threshold
required for radio-loud activity. Over the next few years it will be
extremely interesting to see if high-resolution infrared imaging with 8-m
class telescopes can clarify our picture of high-$z$ quasar hosts
in the same way as has been achieved with the HST at low redshift.


\section*{References}

\noindent
Archibald E., et al. 2001, MNRAS, in press (astro-ph/0002083)\\
Bahcall J.N., Kirhakos S., Schneider D.P., 1994, ApJ, 435, L11\\
Bahcall J.N., et al. 1997, ApJ, 479, 642\\
Blandford R.D., 2000. PTRSA, in press (astro-ph/0001499)\\
Canalizo G., Stockton A., 2000. AJ, 120, 1750\\
Dunlop J.S., et al. 2001, MNRAS, submitted\\
Fabian A.C., 1999, MNRAS, 308, 39\\
Hamilton T.S., Casertano S., Turnshek D.A., 2001, (astro-ph/0011255)\\
Kormendy J., 1977, ApJ, 217, 406\\
Kukula M.J., et al., 2001, MNRAS, in press (astro-ph/0010007)\\
Laor A., 2000, ApJ, 543, L111\\
Magorrian J. \etal, 1998, AJ, 115, 2285\\
McLeod K.K., McLeod B.A., 2001, ApJ, in press (astro-ph/0010127)\\
McLure R.J., Dunlop J.S., 2001, MNRAS, in press (astro-ph/0009406)\\
McLure R.J., et al., 1999, MNRAS, 308, 377\\
Merritt D., Ferrarese L., 2001, MNRAS, 320, L30\\ 
Nolan L.A., et al., 2001, MNRAS, in press (astro-ph/0002027)\\
Percival W.J., et al., 2001, MNRAS in press, astro-ph/0002199\\
Schade D., Boyle B.J., Letawsky M., 2000, MNRAS, 315, 498 \\
Schneider D.P., Gunn J.E., Hoessel J.G., 1983, ApJ, 268, 476\\
Taylor G.L., et al. 1996, MNRAS, 283, 968\\
Wandel A., 1999, ApJ, 519, L39

\end{document}